\documentclass[onecolumn]{aastex631}

\usepackage{amsmath}
\usepackage{graphicx}
\usepackage{amssymb}
\usepackage{mathrsfs}
\usepackage[T1]{fontenc}
\usepackage{natbib}
\usepackage{longtable}
\usepackage{array}
\usepackage{rotating}

\begin{document}

\title{The physical mechanism of radio-quiet turn-on changing-look active galactic nuclei}

\author[0000-0002-7299-4513]{Shuang-Liang Li}
\affiliation{Shanghai Astronomical Observatory, Chinese Academy of Sciences, 80 Nandan Road, Shanghai 200030, People's Republic of China}

\author[0000-0002-2355-3498]{Xinwu Cao}
\affiliation{Institute for Astronomy, School of Physics, Zhejiang University, 866 Yuhangtang Road, Hangzhou 310058, People's Republic of China}

\correspondingauthor{Shuang-Liang Li, Xinwu Cao}
\email{lisl@shao.ac.cn, xwcao@zju.edu.cn}

\begin{abstract}
It is suggested that the variation of mass accretion rate in accretion disk may be responsible for the occurrence of most changing-look active galactic nuclei (CL AGNs). However, the viscous timescale of a thin disk is far longer than the observed timescale of CL AGNs. Though this problem can be resolved by introducing the large-scale magnetic field, the mechanism for radio-quiet CL AGNs with weak/absent large-scale magnetic field remains a mystery. In this work, we assume that the thin accretion disk is collapsed from the inner advection-dominated accretion flow (ADAF) instead of substituting by the outer thin disk through advection. This idea is tested by comparing the cooling timescale ($t_{\rm cool}$) of an ADAF with the observed timescale ($t_{\rm tran}$) of turn-on CL AGNs. We compile a sample of 102 turn-on CL AGNs from the archived data and calculate the cooling timescale of an ADAF with the critical mass accretion rate based on some conventional assumptions. It is found that $t_{\rm cool}$ is much shorter than $t_{\rm tran}$ in most of the CL AGNs, which validates our assumption though $t_{\rm cool}$ is not consistent with $t_{\rm tran}$ ($t_{\rm cool}<t_{\rm tran}$). However, this is reasonable since most of the CL AGNs were observed only two times, indicating that the observed timescale $t_{\rm tran}$ is the maximum value because the changing-look can indeed happen before the second observation.
\end{abstract}



\section{Introduction}
Depending on whether the broad emission line region (BLR) is obscured by a torus, active galactic nuclei (AGNs) can be roughly classified into type 1 and type 2 based on the presence or absence of broad emission lines in the AGN unified model (e.g., \citealt{seyfert_nuclear_1943,antonucci_unified_1993}). Usually, the type of AGN will remain unchanged within the viscous timescale ($\sim 10^4-10^7$ years) of a thin accretion disk in AGNs \citep{1973A&A....24..337S,2002apa..book.....F}. However, some AGNs (several hundreds) have been observed to change their type in much shorter timescale ranging from several months to tens of years (e.g., \citealt{parker_detection_2016,gezari_iptf_2017,yang_discovery_2018,macleod_changing-look_2019,wang_x-ray_2020,2022ApJ...933..180G,2024arXiv240807335D,2024ApJS..272...13P,2025ApJ...980...91Y}). These objects are named as changing-look AGNs (CL AGNs).  Exploring the physics of CL AGNs can help us understand the accretion process in AGNs. 

When CL AGNs are first discovered, people consider that the obscuration of torus on BLR should be responsible for the occurrence of CL AGNs (e.g., \citealt{goodrich_spectropolarimetry_1989, tran_extreme_1992,elitzur_unification_2012}). However, if this scenario is correct, the variability of infrared emissions should be very weak, inconsistent with the large amplitude variations in CL AGNs \citep{2017ApJ...846L...7Sheng}. 
The low polarization of optical emissions in CL AGNs provides further evidence opposing the torus obscured model \citep{hutsemekers_polarization_2019}. The other popular model for CL AGNs is the change of global mass accretion rate in disk, which is the most promising mechanism so far and has been supported by the co-variation of multi-wavelength emissions (e.g., \citealt{ross_new_2018,2020A&A...641A.167S,2023ApJ...953...61Y}). However, as mentioned above, the viscous timescale ($t_{\rm vis}\sim R/V_{\rm R}$) of a thin disk is far longer than the observed timescale of CL AGNs. Obviously, by constraining the disk size $R$ or/and enhancing the radial velocity $V_{\rm R}$ can help to solve this inconsistency. \citet{2020A&A...641A.167S} suggested that the instability of a narrow disk region between an inner advection-dominated accretion flows (ADAF) and an outer thin disk can qualitatively explain the CL AGNs with multi-CL (see also \citealt{2021ApJ...910...97P}). It is found that the radial velocity of a thin disk threaded by the large-scale poloidal magnetic field can be significantly improved for the reason that the outflow driven by the magnetic field can transfer a large fraction of the angular momentum of accretion disk. (e.g, \citealt{2013ApJ...765..149Cao,2014ApJ...786....6L, 2021ApJ...916...61F,2023ApJ...958..146W}). In addition, the disk scale-height elevated by the large-scale toroidal field can also improve the radial velocity (e.g, \citealt{2019MNRAS.483L..17Dexter}). All these models can qualitatively fix the inconsistency between the observed timescale and the theoretical viscous timescale. However, most of the CL AGNs are radio-quiet, i.e., the large-scale magnetic field threading on the accretion disk should be very weak or absent. In this circumstance, the radial velocity of thin disk cannot be significantly improved, leading to the much longer viscous timescale of a thin disk than the typical observed timescale of CL AGNs. Therefore, another mechanism is necessary to explain the the observed timescale in radio-quiet CL AGNs.

On the other hand, the accretion physics of black hole should be scaled-free. It is well known that the accretion flow of low-hard state in black hole X-ray binaries is composed by an inner ADAF and an outer thin accretion disk, while the high-soft sate is dominated by a thin disk only (e.g., \citealt{1997ApJ...489..865E, 2014ARA&A..52..529Y}). Some works argued that CL AGNs have analogous accretion structure with the state transition of black hole X-ray binaries, where the type 2 and type 1 of CL AGNs correspond to the low-hard and high-soft state of X-ray binaries, respectively (e.g., \citealt{2019ApJ...883...76R,2022ApJ...927..227L}). The reasons include the similar relationship between their UV–to-X-ray spectral index $\alpha_{\rm OX}$ and the Eddington ratio $\lambda$ and the similar critical Eddington ratio to separate the low/high state in X-ray binaries and that to separate the type 1/2 in CL AGNs. Inspired by this opinion, we consider the possibility that the inner ADAF collapses to a thin accretion disk through radiative cooling, instead of replacing by the outer thin disk through radial advection. This mechanism will resolve the problem that large-scale magnetic filed may be very weak or absent in radio-quiet AGNs. In this work, we investigate the cooling timescale of an ADAF during the state transition and compare it with the observed timescale of turn-on CL AGNs.

\section{Model}\label{model}

In the low-hard state of black hole X-ray binaries, the X-ray emissions consist of a strong power-law component plus a weak blackbody component, which is believed to come from an inner optically thin and geometrically thick ADAF and an outer optically thick and geometrically thin accretion disk, respectively \citep{2006ARA&A..44...49R,2014ARA&A..52..529Y}. The radiative efficiency of an ADAF is roughly proportional to the square of mass accretion rate, resulting in the more efficient cooling rate than the heating rate when the Eddington ratio increases ($\eta\propto \dot{m}^2$, \citealt{1994ApJ...428L..13N,1995ApJ...452..710N}). Therefore, an ADAF will collapse to a thin accretion disk and the binary system will enter the high-soft state at a critical Eddington ratio ($\lambda_{\rm Edd}\gtrsim 0.01$).

The energy equation of an ADAF can be given as  
\begin{equation}
q^+=q^-+q_{\rm adv}=q^-+f_{\rm adv}q^+, \label{energy}
\end{equation} 
where $q^+$, $q^-$, $q_{\rm adv}$ are the viscously heating rate, the cooling rate and the advected rate of energy, respectively. In a standard ADAF, $q^+\gg q^-$, so $f_{\rm adv}$ is $\sim 1$ \citep{1994ApJ...428L..13N,1995ApJ...452..710N}. However, the Coulomb coupling between electron and proton will become more and more efficiently as the Eddington ratio increases. When closing to a critical value, a large fraction of the released gravitational energy will be taken away through radiation, leading to the formation of a thin disk \citep{1995ApJ...452..710N,1998tbha.conf..148N}. 

As \citet{1995ApJ...452..710N}, the cooling timescale of an ADAF can be written as
\begin{equation}
    t_{\rm cool}\sim \frac{u}{q^{-}}, \label{tcool}
    \end{equation} 
where $u\sim3n_{\rm i}kT_{\rm i}/2\sim 3P/2$ is the internal energy of gas, and $P\cong P_{\rm i}$ is the pressure of gas. The viscous dissipation rate of energy $q^+$ is
\begin{equation}
  q^+=-\alpha PR\frac{d\Omega}{dR}  \cong \frac{3}{2} f_{\Omega}\alpha P \Omega_{\rm K}, \label{heating}
\end{equation} 
where $\alpha$ is the dimensionless viscosity parameter. Parameter $f_{\Omega}$ is given by $f_{\Omega}=\Omega/\Omega_{\rm K}$, where $\Omega$ and $\Omega_{\rm K}$ are the angular velocity and the Keplerian angular velocity of the gas, respectively. Combining with equations (\ref{energy}) and (\ref{heating}), equation (\ref{tcool}) can be rewritten as 
\begin{equation}
    t_{\rm cool}=\frac{1}{(1-f_{\rm adv})f_{\Omega}\alpha \Omega_{\rm K}}. \label{tcool2}
\end{equation}
As mentioned above, the advected rate of energy ($q_{\rm adv}=f_{\rm adv}q^+$) will decrease with increasing Eddington ratio. However, the detailed value of $f_{\rm adv}$ is unclear, which will vary with the viscosity parameter $\alpha$ and the ratio of gas pressure to the magnetic pressure $\beta$ \citep{1995ApJ...452..710N,1998tbha.conf..148N}. In this work, we adopt a typical value of $f_{\rm adv} \sim 0.3$ when the critical mass accretion rate $\dot{m}_{\rm crit}$ is adopted, where $\dot{m}_{\rm crit}$ is given as $\dot{m}_{\rm crit}\cong \alpha^2$ in theory (see e.g., \citealt{1995ApJ...452..710N}). The angular velocity of a thin accretion disk is usually adopted to be Keplerian because its disk temperature is low. However, the temperature of an ADAF is very high (the ion temperature can be close to the virial temperature) since most of the released gravitational energy is stored in the gas instead of being radiated away.  The higher temperature will produce a strong pressure gradient force, which offsets part of the gravitational force and leads to a sub-Keplerian angular velocity (see e.g., \citealt{1994ApJ...428L..13N,1995ApJ...452..710N}).  Therefore, the angular velocity of an ADAF before collapsing to a thin disk is sub-Keplerian. However, the angular velocity will increase with the decrease of temperature when the mass accretion rate is close to the critical mass accretion rate. So we adopt 90\% of the Keplerian velocity ($f_{\Omega}=0.9$) in this work. Our results will be qualitatively the same  even a fully Keplerian velocity as in a thin disk is adopted (resulting in a  $t_{\rm cool}$ value 10\% smaller).

The critical mass accretion rate $\dot{m}_{\rm crit}$ ($=\dot{M}_{\rm crit}/\dot{M}_{\rm Edd}$) is equal to $\lambda_{\rm crit}$ assuming a conventional radiation efficiency $\eta=0.1$, where $\dot{M}_{\rm Edd}=L_{\rm Edd}/0.1 c^2$ is the Eddington accretion rate. However, the detailed value of $\dot{m}_{\rm crit}$ in CL AGNs are unknown because most of objects only have two spectroscopic observations, i.e., one faint state and one bright state, respectively.  Therefore, we adopt the value of Eddington ratio $\lambda_{\rm Edd}$ at bright state as $\dot{m}_{\rm crit}$ (see column 3 in table 1). In addition, the bolometric luminosity of CL AGNs in our sample is estimated with the continuum flux at 5100\AA, indicating the temperature at the transition radius $R_{\rm tr}$ between the inner ADAF and outer thin disk should be larger than the temperature corresponding with wavelength 5100\AA ($\sim 5700 K$). Therefore, $R_{\rm tr}$ can be estimated by solving the following equation:
\begin{equation}
 \sigma T_{\rm eff}^4=\frac{3GM_{\rm bh}\dot{M}}{8\pi R^3}\left(1-\sqrt{\frac{6R_{\rm g}}{R}}\right),
\end{equation}
where $\sigma$, $G$, and $R_{\rm g}=GM_{\rm bh}/c^2$  are the  Stefan-Boltzmann constant, the gravitational constant and the gravitational radius, respectively. With $f_{\rm adv}$, $f_{\Omega}$, $\alpha$, and the transition radius $R_{\rm tr}$, the cooling timescale of an ADAF at $R_{\rm tr}$ can be gotten with equation (\ref{tcool2}).

\section{Sample}\label{sample}

Most of CL AGNs discovered in early years are identified as turn-off CL AGNs because they are usually picked up firstly from the spectroscopically-confirmed quasars 
\citep{2025ApJ...980...91Y}. It is hard to find a relatively large turn-on CL AGN sample until this year. In order to compare with the theoretical model, the black hole mass, Eddington ratio and timescale of CL AGNs are all required (see Section \ref{results} for details). We compile a sample of 102 turn-on CL AGNs by searching the archived data in this work (see table 1). Most of the objects (82) are picked up from \citet{2025ApJ...980...91Y}, while other 12 and the last 8 objects are selected from \citet{2024ApJS..272...13P} and \citet{2024arXiv240807335D}, respectively. To ensure the formal consistency, we rewrite all the objects as SDSSJXXXX+/-XXXX in column (1). The black hole mass $m_{\rm bh}$ in column (3) are all estimated with the width of broad emission lines \citep{2006ApJ...641..689V}. Column (4) gives the Eddington ratio $\lambda_{\rm Edd}$ of bright states in CL AGNs. Column (5) and (6) represent the observed modified Julian dates of faint state and bright state, respectively. The transition timescale $t_{\rm tran}$ in column (7) and cooling timescale $t_{\rm cool}$ in column (9) indicate the observed timescale of CL AGNs and the cooling timescale of ADAF given by theory, respectively. $t_{\rm tran}$ has been corrected to the value in rest frame of the object ($t_{\rm tran}=(\rm{MJD}_2-\rm{MJD}_1)/(1+z)$, where $z$ is the redshift).

\newpage
\begin{longtable*}{l}
\caption{The sample.} \\
\hspace{-3.8cm}
\resizebox{1.2\textwidth}{!}{
\begin{tabular}{cccccclcl|cccccclcl}
\hline
{Name} & {z} & {$m_{\rm bh}$} &  {$\lambda_{\rm Edd}$} & {$\rm{MJD}_1$} & {$\rm{MJD}_2$} & {$t_{\rm tran}$} & {Ref.} & {$t_{\rm cool}$} &{Name} & {z} & {$m_{\rm bh}$} &  {$\lambda_{\rm Edd}$} & {$\rm{MJD}_1$} & {$\rm{MJD}_2$} & {$t_{\rm tran}$} & {Ref.} & {$t_{\rm cool}$}   \\ 
{(1)} &  {(2)} &  {(3)} &  {(4)} &
 {(5)} &  {(6)} &  {(7)} &
 {(8)} &  {(9)} & {(1)} &  {(2)} &  {(3)} &  {(4)} &
 {(5)} &  {(6)} &  {(7)} &
 {(8)} &  {(9)}\\
\hline
SDSSJ0002-0027  & 0.291  &  	8.94${\pm0.02}$  &  -1.99   & 52559 
 &  55477 & 	2260   &  	1  &  	2764${\pm63}$    	&
SDSSJ0023+0035  & 0.422     &  		9.17${\pm0.04}$  &  -1.61   & 51816
 &  55480      &  		2577   &  		1  &  		3507${\pm162}$ \\
SDSSJ0040+1609  &  0.294 &		7.85${\pm0.04}$  &  		-1.49   & 51884 & 60181   &		6412  &  		2  &  		813${\pm38}$  &  
SDSSJ0047+1541  & 0.031 & 	7.07${\pm0.21}$  &   	-2.22  & 51879 & 59856  &	7734  &  	2  &  	322${\pm87}$ \\
SDSSJ0107+2428   & 0.160 &  	8.37${\pm0.05}$  &    	-2.07  &  57367 &  59970 & 	2244  &    	2  &   	1479${\pm85}$   &  
SDSSJ0110+0026    &  0.019 &	7.09${\pm0.03}$   &   	-2.12   &   51794 
& 59880 &	7937  &   	2   &  	329${\pm11}$ \\
SDSSJ0132+1501   &  0.170 &		8.15${\pm0.02}$   &  	 	-2.21   &  	51884  &  60240  & 	7142  &  	 	2   &  		1101${\pm25}$   &  	 
SDSSJ0141+0105   & 0.101 & 		8.39${\pm0.08}$   &  	 	-2.07   &  51788 & 58794 &	 	6363   &  	 	2   &  		1512${\pm142}$ \\
SDSSJ0144+3140    &  0.124  &		7.66${\pm0.03}$   &  	 	-1.48   &  	57725  &  58133  & 	363  &  	 	3   &  		663${\pm23}$   &  
SDSSJ0146+1311   &  0.160  &		8.10${\pm0.02}$    &  		-2.07   &  	51820  &  60196  & 	7221 &  	 	2   &  		1092${\pm25}$ \\
SDSSJ0225+0030   &  0.504  &		9.27${\pm0.09}$   &  	 	-2.15    &  55445  & 55827  &		254 &  		1   &  		4112${\pm431}$    &  	
SDSSJ0334+0051   & 0.429  & 		8.39${\pm0.05}$ 
    &  		-1.65   &  	52178 & 60193  & 	5609   &  		2   &  		1489${\pm86}$ \\
SDSSJ0803+2207   &  0.125  &		7.59${\pm0.02}$   &  	 	-1.72   &  
 52943 &	 60322 & 	6559   &  	 	2   &  		595${\pm14}$   &  	 
SDSSJ0813+4608   &  0.054  &		7.37${\pm0.02}$   &  	 	-1.53   &  51877 &	60322 & 	8014   &  	 	2   &  		477${\pm11}$\\ 
SDSSJ0819+3019   &  0.098  &		7.98${\pm0.01}$   &  	 	-1.65   &  	52619 & 60239 & 	6943   &  	 	2   &  		937${\pm11}$   &  
SDSSJ0823+4202   &  0.126  &		7.56${\pm0.06}$   &  	 	-1.65   &  	52266  & 60016  & 	6883   &  	 	2   &  		583${\pm41}$\\ 
SDSSJ0829+2319   &  0.144 &		8.09${\pm0.01}$   &  	 	-2.69   &  	 53349 &  60323 &	6096   &  	 	2   &  		934${\pm11}$   &  	 
SDSSJ0829+4154   &  0.126 & 		8.71${\pm0.02}$   &  	 	-2.07   &  	52266 & 54524 &	2005   &  	 	1   &  		2162${\pm50}$   \\ 	 
SDSSJ0831+3646  & 0.195 & 	8.26${\pm0.09}$   &  	 	-1.74   &  	52312 
&  57367 & 	4230  &  	 	3   &  		1335${\pm142}$   &  
SDSSJ0837+0356   &  0.064 & 		7.13${\pm0.05}$   &  	 	-2.07   &  52646 &  59288  & 6245   &  	 	2   &  		365${\pm21}$    \\  	
SDSSJ0838+3719   &  0.211  &		8.72${\pm0.04}$   &  	 	-1.77   &  	52320 & 59200  & 	5681   &  	 	3   &  		2160${\pm98}$  &  
SDSSJ0853+2128   &  0.084 &		7.83${\pm0.03}$   &  	 	-1.56   &  	 53680 & 59078 &	4979   &  	 	3   &  		776${\pm27}$    \\  	
SDSSJ0854+1113   &  0.167 &		8.04${\pm0.09}$   &  	 	-1.76   &  	 54084 & 58794 &	4036   &  	 	2   &  		1017${\pm109}$  &  
SDSSJ0901+2907   &  0.218 &		8.71${\pm0.02}$   &  	 	-2.59   &  	 56341 & 60239 &	3200   &  	 	2   &  		2094${\pm45}$    \\ 	
SDSSJ0906+4046   &  0.167 &		8.07${\pm0.05}$   &  	 	-2.85   &  	 52668 & 60323 &	6560   &  		2   &  		944${\pm53}$   &  
SDSSJ0908+0755   &  0.144 &		8.68${\pm0.02}$    &  		-2.93   &  	 52973 & 60240 &	6352   &  	 	2   &  		2245${\pm48}$   \\  	 
SDSSJ0910+1907   &  0.188 &		7.87${\pm0.03}$   &  	 	-1.91    &  	53700 & 60029 &	5327    &  	 	2   &  		832${\pm29}$  &  
SDSSJ0914+0502   &  0.143 &		8.49${\pm0.08}$   &  	 	-2.92    &  52652 & 58879 &	5448    &  		2   &  		1844${\pm169}$    \\  	
SDSSJ0915+4814   &  0.101 &		8.07${\pm0.02}$   &  	 	-2.71   &  	 52637 & 59501 &	6234   &  	 	2   &  		1114${\pm26}$   &  
SDSSJ0926-0006   &  0.201 &		6.95${\pm0.04}$    &  		-1.11    &  52000 & 59227 &		6017    &  		2   &  		293${\pm14}$    \\  	
SDSSJ0936+2726   &  0.103 &		6.68${\pm0.02}$    &  		-0.79   &  	 53385 & 59527 &	5568    &  	 	2   &  		218${\pm5}$   &  
SDSSJ0937+2602   &  0.162 &		7.58${\pm0.02}$    &  		-1.19 	   &  53733 & 57369 &	3129    &  		3   &  		612${\pm14}$    \\  	
SDSSJ0947+5449   &  0.620 &		8.35${\pm0.10}$    &  		-1.09    &  56011 & 58794 &		1718    &  		2   &  		1471${\pm176}$   &  
SDSSJ0951+3416   & 0.132  & 		7.85${\pm0.01}$   &  	 	-1.74   &  53388 & 60240 &	 	6053   &  	 	2   &  		840${\pm10}$    \\  	
SDSSJ0952+2229   &  0.081 &		6.44${\pm0.06}$    &  		-1.28    &  53734 & 60322 &		6094    &  		2	   &  	163${\pm12}$   &  
SDSSJ0955+1037   &  0.284 &		8.39${\pm0.04}$   &  	 	-0.93   &  	 52996 & 59200 &	4832   &  	 	3   &  		1529${\pm71}$    \\ 	
SDSSJ1000+0354   &  0.215 &		7.26${\pm0.11}$   &  		-1.22   &  	 52289 & 59333 &	5798   &  	 	2   &  		428${\pm57}$   &  
SDSSJ1002+0303   &  0.023 &		7.18${\pm0.03}$    &  		-1.79    &  52235 & 59250 &		6855   &  	 	2   &  		371${\pm13}$  \\ 	 
SDSSJ1003+0202   & 0.125 & 		7.74${\pm0.41}$   &  	 	-2.53    &  52235 & 58494 &		5564    &  		2   &  		763${\pm444}$   &  
SDSSJ1003+0458   &  0.124 &		8.02${\pm0.14}$   &  	 	-2.55   &  	 52325 & 58495 &	5489   &  	 	2	   &  	1019${\pm172}$  \\  	
SDSSJ1003+3525   &  0.119  &		8.28${\pm0.05}$   &  	 	-2.01    &  53389 & 58612 &		4668    &  		2	   &  	1291${\pm74}$   &  
SDSSJ1003+1932   &  0.467 &		8.84${\pm0.03}$    &  		-1.58    &  53762 & 57817 &		2764    &  		1	   &  	2514${\pm84}$    \\ 	
SDSSJ1009+3539	   & 0.110 & 	8.35${\pm0.07}$    &  		-2.17   &  	 53389 & 58896 &	4961   &  	 	2   &  		1446${\pm118}$   &  
SDSSJ1011+5442   &  0.247 &		7.63${\pm0.01}$    &  		-0.79    &  57073 & 60291 &		2581    &  	 	2   &  		645${\pm7}$   \\  	
SDSSJ1012+5559   & 0.126 & 		7.71${\pm0.11}$   &  	 	-1.61    &  52407 & 58163 &		5112    &  		2   &  		679${\pm90}$  &   
SDSSJ1017+0658   &  0.045 &		7.63${\pm0.11}$    &  		-2.19   &  	 55679 & 58843 &	3027    &  		2   &  		628${\pm83}$    \\  	
SDSSJ1020+2437   &  0.189 &		9.13${\pm0.01}$   &  		-2.39    &  53734 & 60240 &		5472   &  	 	2   &  		3523${\pm37}$   &  
SDSSJ1058+6338   &  0.202 &		7.42${\pm0.05}$    &  		-1.39    &  54498 &	60059 & 	4626    &  		2   &  		505${\pm30}$  \\  	
SDSSJ1102+5722   &  0.142 &		8.21${\pm0.05}$    &  		-2.24    &  52427 & 60012 &		6642    &  		2   &  		1137${\pm66}$   &  
SDSSJ1105+0516   &  0.091 &		7.67${\pm0.08}$   &  	 	-1.86   &  	 52353 & 59231 &	6303   &  	 	2   &  		645${\pm61}$    \\  	
SDSSJ1110+0804   &  0.116 &		7.51${\pm0.17}$    &  		-2.48    &  52723 & 58136 &		4850   &  	 	2	   &  	520${\pm110}$   &  
SDSSJ1111+1401   &  0.110 &		7.48${\pm0.07}$   &  	 	-1.55    &  53377 & 58612 &		4716    &  		2	   &  	528${\pm44}$   \\ 	
SDSSJ1115+0544   & 0.090 & 		7.63${\pm0.06}$    &  		-1.61   &  	 52326 & 57393 &	4649    &  		2   &  		620${\pm44}$   &  
SDSSJ1126+0510	   & 0.170 &  	8.22${\pm0.04}$    &  		-2.29    &  52376 & 60059 &		6567    &  		2   &  		1239${\pm57}$   \\  	
SDSSJ1132+0357   &  0.091 &		7.57${\pm0.04}$    &  		-1.80    &  52642 & 58612 &		5472    &  		2   &  		571${\pm26}$   &  
SDSSJ1136+4450   &  0.116 &		8.30${\pm0.02}$    &  		-1.47    &  53083 & 57732 &		4166    &  		1   &  		1385${\pm31}$   \\  	
SDSSJ1149+3351   &  0.081 &		8.12${\pm0.02}$    &  		-2.18    &  53491 & 60052 &		6070    &  		2	   &  	1104${\pm25}$   &  
SDSSJ1155+0048   &  0.209 & 	9.21${\pm0.01}$    &  		-2.82    &  51662 & 59317 &		6332    &  		2   &  		3430${\pm36}$  \\  	
SDSSJ1158+1003   &  0.070 &		7.47${\pm0.05}$    &  		-3.38    &  52733 & 58844 &		5712   &  	 	2   &  		515${\pm29}$  &  
SDSSJ1159+0513   & 0.059 & 		7.73${\pm0.01}$   &  	 	-1.71   &  	 52375 & 60319 & 	7499    &  		2	   &  	706${\pm8}$   \\  	
SDSSJ1159+6545   &  0.122 &		8.56${\pm0.07}$   &  	 	-2.08    &  52316 & 58496 &		5508   &  	 	2	   &  	1807${\pm147}$   &  
SDSSJ1205+3335   & 0.299 & 		8.22${\pm0.02}$    &  		-1.06    &  56418 & 58467 &	1577   &  	 	3   &  		1227${\pm28}$    \\  	
SDSSJ1226+4559	   & 0.109 & 	7.87${\pm0.07}$   &  	 	-2.81   &  	 52821 & 60013 &	6485    &  		2   &  		792${\pm64}$   &  
SDSSJ1231+3232   &  0.065 &		7.76${\pm0.02}$    &  		-1.71   &  	 53472 & 60013 &	6139    &  		2   &  		730${\pm17}$   \\  	
SDSSJ1239+0739   &  0.133 &		7.48${\pm0.09}$   &  	 	-1.87    &  53474 & 60060 &		5813    &  		2   &  		514${\pm55}$   &  
SDSSJ1311+0705  & 0.301 & 	7.79${\pm0.19}$    &  		-1.38   &  	 	54507 & 59295 &  3680    &  		2   &  		778${\pm123}$   \\ 	
SDSSJ1311+4300	   & 0.285 &  	7.91${\pm0.15}$    &  		-1.44    &  57783 & 60013 & 		1735   &  	 	2   &  		874${\pm162}$   &  
SDSSJ1327+4025   &  0.078 &		7.75${\pm0.06}$   &  	 	-2.75    &  53091 &	59250 &	5715    &  		2   &  		743${\pm52}$   \\  	
SDSSJ1328+6227   &  0.091 &		7.81${\pm0.04}$    &  		-2.63    &  52339 & 58197 &		5371    &  		2	   &  	733${\pm33}$   &  
SDSSJ1340+4006	   &  0.171 &	8.22${\pm0.02}$    &  		-1.89    &  53050 & 59729 &		5704    &  		2   &  		1264${\pm29}$   \\  	
SDSSJ1341-0049   &  0.175 &		7.31${\pm0.25}$    &  		-1.40    &  51671 & 58256 &		5604   &  	 	2   &  		441${\pm145}$   &  
SDSSJ1413+5305   & 0.456 & 		8.38${\pm0.03}$    &  		-0.62    &  52762 & 58289 &		3796   &  	 	1   &  		1506${\pm52}$  \\  	
SDSSJ1433+4943   & 0.125 & 		7.97${\pm0.13}$    &  		-2.36    &  52460 & 59317 &		6095    &  		2   &  		861${\pm136}$   &  
SDSSJ1442+5558	   & 0.077 & 	7.24${\pm0.03}$    &  		-1.36    &  52668 & 59727 &		6554    &  		2   &  		406${\pm14}$    \\  	
SDSSJ1511+2101   &  0.081 &		8.15${\pm0.03}$   &  	 	-1.40   &  	 54525 & 59732 &	4819    &  	 	3	   &  	1142${\pm39}$ 	  &  
SDSSJ1524+4327   &  0.198 &		8.34${\pm0.05}$    &  		-2.41    &  52468 & 58987 &		5442    &  		2   &  		1433${\pm83}$   \\
SDSSJ1538+4607   & 0.203 & 		8.49${\pm0.03}$    &  		-2.63   &  	 52781 & 60195 &	6163    &  		2   &  		1565${\pm52}$  	  &  
SDSSJ1552+2102   &  0.172 &		7.91${\pm0.02}$   &  	 	-1.51    &  53557 & 60193 & 		5662    &  		2   &  		850${\pm19}$ \\
SDSSJ1611+1642   &  0.155 &		8.15${\pm0.01}$   &  	 	-1.89    &  53555 & 60206 &	5758    &  		2   &  		1168${\pm13}$   &    	
SDSSJ1612+3113   & 0.177 & 		8.95${\pm0.02}$   &  	 	-3.16    &  58244 & 60107 &		1583    &  		2	   &  	2280${\pm44}$ \\
SDSSJ1612+3816   &  0.199 &		8.05${\pm0.03}$    &  		-2.22    &  53146 & 60117 &		5814    &  		2	   &  	973${\pm34}$   &    	
SDSSJ1636+3930   &  0.182 &		7.60${\pm0.08}$   &  	 	-1.52   &  	 52381 & 60206 & 	6620   &  	 	2   &  		626${\pm60}$   \\ 
SDSSJ1638+4712   &  0.114 &		7.97${\pm0.04}$    &  		-2.32   &  	 52144 & 59726 &	6806   &  	 	2   &  		904${\pm41}$     &   	
SDSSJ1640+2819   & 0.181 & 		7.93${\pm0.05}$   &  	 	-2.05   &  	 53472 & 59724 &	5294   &  		2   &  		923${\pm53}$   \\
SDSSJ1642+4423   &  0.227 &		7.97${\pm0.04}$    &  		-1.58    &  53557 & 60193 &	5408   &  	 	2   &  		943${\pm43}$     &    	
SDSSJ1643+5000   &  0.060 &		7.39${\pm0.03}$   &  	 	-2.47    &  57187 & 60206 &		2848   &  	 	2   &  		460${\pm16}$ \\
SDSSJ1652+3416	   &  0.263 &	8.18${\pm0.05}$   &  	 	-1.93    &  52435 & 59314 &	5447   &  	 	2   &  		1152${\pm67}$     &   	
SDSSJ1723+5504   &  0.295 &		8.80${\pm0.15}$    &  		-1.73 	   &  51813 & 51997 &	142    &  		1	   &  	2300${\pm420}$  \\
SDSSJ2120-0056   &  0.251 &		8.38${\pm0.03}$    &  		-2.02    &  52932 & 60181 &	5795    &  		2   &  		1428${\pm49}$      &   	
SDSSJ2150-0106	   &  0.088 &	7.45${\pm0.08}$    &  		-2.10    &  53172 & 58794 &		5168   &  	 	2   &  		507${\pm48}$   \\
SDSSJ2156+2907   &  0.092 &		7.49${\pm0.10}$   &  	 	-1.52   &  	 56101 & 59879 &	3460   &  	 	2	   &  	553${\pm66}$      &   	
SDSSJ2159+0556   &  0.111 &		8.15${\pm0.02}$   &  	 	-2.81    &  58431 & 60197 &		1590    &  		2   &  		1081${\pm25}$  \\
SDSSJ2201+0352   &  0.136 &		7.60${\pm0.07}$   &  	 	-2.05   &  	 58429 & 60197 &	1556    &  		2   &  		636${\pm52}$      &   	
SDSSJ2203+1124   &  0.186 &		6.82${\pm0.09}$    &  		-1.29    &  52224 & 58105 &		4959    &  		2	   &  	249${\pm27}$   \\
SDSSJ2225+2019   &  0.155 &		8.18${\pm0.03}$    &  		-2.72    &  56189 & 60197 &	3470    &  		2	   &  	1245${\pm42}$     &   	
SDSSJ2146+0009   &  0.621 &		9.44${\pm0.02}$    &  		-2.21    &  52968 & 55478 &		1548    &  		1   &  		4618${\pm95}$   \\
SDSSJ2241-0121   &  0.058 &		8.31${\pm0.03}$   &  	 	-2.23    &  55824 & 57724 &	1796   &  	 	1	   &  	1288${\pm44}$      &    	
SDSSJ2252+0109   & 0.534 & 		8.83${\pm0.03}$   &  	 	-1.42   &  	 52178 & 55500 &	2166    &  		1   &  		2528${\pm87}$   \\
SDSSJ2320+1448   &  0.082 &		7.60${\pm0.04}$    &  		-2.36    &  52258 & 59501 &	6693    &  		2   &  		569${\pm26}$      &    	
SDSSJ2336+1514   &  0.147 &		7.73${\pm0.02}$    &  		-1.48    &  52234 & 59821 &		6615    &  		2   &  		718${\pm16}$   \\
SDSSJ2346+0109   &  0.509 &		9.00${\pm0.19}$   &  	 	-1.64    &  52524 & 57684 &	3419    &  		1	   &  	2986${\pm708}$      &    	
SDSSJ2348+0741   &  0.464 &		7.87${\pm0.06}$    &  		-1.34 	   &  56164 & 60240 &	2784  &  	 	2   &  		850${\pm60}$  \\

\hline
\multicolumn{18}{p{25cm}} {Notes: Col. (1): Source name. Col. (2): Redshift. Col. (3): Black hole mass $m_{\rm bh}=\log(M_{\rm BH}/M_{\odot})$, where $M_{\rm BH}$ and $M_{\odot}$ are the black hole mass and solar mass, respectively. Col. (4): Eddington ratio of CL AGNs in bright state. Col. (5): modified Julian date of faint state. Col. (6): modified Julian date of bright state. Col. (7): Observational timescale fo CL AGNs. Col.(8): References. Col. (9): Cooling timescale in theory with uncertainty derived from the uncertainty of black hole mass. } \\
\multicolumn{18}{p{25cm}}{References. 1, \citet{2024ApJS..272...13P}; 2, \citet{2025ApJ...980...91Y}; 3, \citet{2024arXiv240807335D}. }
\end{tabular}
}

\end{longtable*}

\section{Results}\label{results}

\begin{figure}
\centering
\includegraphics[width=14cm]{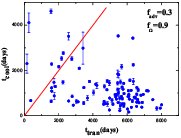}
\caption{Comparison of the observed timescale $t_{\rm tran}$ of CL AGNs with the cooling timescale $t_{\rm cool}$ of an ADAF, where the solid red line represent $t_{\rm tran}=t_{\rm cool}$.} \label{f1}
\end{figure}

\begin{figure}
\centering
\includegraphics[width=14cm]{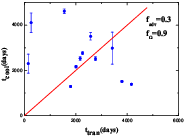}
\caption{Same as figure \ref{f1}, but for the objects picked up from \citet{2024ApJS..272...13P} only.} \label{f2}
\end{figure}

We compare the observed timescale $t_{\rm tran}$ of CL AGNs and the cooling timescale $t_{\rm cool}$ of an ADAF in Figure \ref{f1}, where the red solid line represents $t_{\rm tran}=t_{\rm cool}$. It is found that $t_{\rm cool}$ is much shorter than $t_{\rm tran}$ in most of CL AGNs, while $t_{\rm cool}\gg t_{\rm tran}$ is found in two CL AGNs, i.e., SDSSJ0225+0030 and SDSSJ1723+5504. This result confirms our assumption that the inner ADAF can collapse to a thin disk within the observed timescale of turn-on CL AGNs. The origin for the two special objects with $t_{\rm cool}\gg t_{\rm tran}$ may be related with the large-scale magnetic field (see section \ref{summary} for details).

\section{Summary and Discussion} \label{summary}

For most of radio-quiet CL AGNs where the large-scale magnetic field is absent or very weak, the viscous timescale of a thin disk would be too long compered with the observed timescale. Therefore, it is possible that, instead of being replaced by the outer thin disk, the formation of inner thin disk can be collapsed from the inner ADAF in CL AGNs. We compile a sample of 102 turn-on CL AGNs from the archived data to test this idea in this work. It is found that the the cooling timescale $t_{\rm cool}$ of an ADFA is shorter than the observed timescale $t_{\rm tran}$ of most turn-on CL AGNs, supporting our assumption though they are not consistent in most of objects. This inconsistency between $t_{\rm cool}$ and $t_{\rm tran}$ may originate from: 1) the observed timescale $t_{\rm tran}$ is the upper limit instead of detailed value because the bright state in most of CL AGNs is observed one time only.
Even the light-curves in these objects are available, however, the variation of Eddington ratio during the transition is very small. People can't detect obvious flares from the light-curves (see e.g., \citealt{2025ApJ...980...91Y}). CL AGNs can be identify only from the dis/appearance of broad emission lines through spectroscopic observations; 2) adopting the Eddington ratio at bright state as the critical mass accretion rate is also an upper limit, resulting on a larger viscosity parameter $\alpha$ and smaller $t_{\rm cool}$. Furthermore, it is found that a large fraction of objects picked up from \citet{2024ApJS..272...13P} have more than two spectroscopic data. So we compare $t_{\rm cool}$ with $t_{\rm tran}$ individually for the objects selected from \citet{2024ApJS..272...13P} and find the result is much better for this subsample (see Figure \ref{f2}). Therefore, the main caveat of this manuscript is that most of the objects only have two spectral observations, leading to a much longer observed timescales ($t_{\rm tran}$) than the actual timescales of CL. If better spectral data can be gathered by shortening observation intervals and increasing the number of observations in the future, we can measure $t_{\rm tran}$ more accurately and further test our model in this manuscript. 

Different with most other objects, SDSSJ0225+0030 and SDSSJ1723+5504 have much shorter observed timescales, which may be due to the presence of large-scale magnetic field. The large-scale magnetic filed threading on the outer thin accretion disk can launch strong outflows and take away the angular momentum of the disk, leading to a much faster radial velocity (see, e.g., \citealt{2013ApJ...765..149Cao,2014ApJ...786....6L}). Therefore, the inner ADAF may be dragged inwards to the central black hole before it is totally cooling to a thin disk.

We investigate the cooling timescale of an ADAF and compare it with the turn-on CL AGNs only in this work. For turn-on CL AGNs, we can only focus on the collapse of an ADAF and the formation of a thin disk through radiative cooling. While for AGNs in bright state, the accretion process can be well described by a disk-corona model \citep{2017A&A...602A..79L,2018MNRAS.477..210Q}, where the optical-UV flux is emitted from a thin accretion disk and the hard X-ray flux comes from the inverse Compton scattering of optical-UV photons by the hot corona above the thin disk. When CL AGNs are turn-off, we have to consider simultaneously both the process of gas evaporation in thin disk and the process of gas condensation in corona, while these physical processes are still unclear so far. We will further explore this issue in future.

\section* {Acknowledgements}
We thank the reviewer for his/her valuable comments, which help to clarify this manuscript. SLL thank Qian Yang for useful comments and sharing their data before official publication. This work is supported by the National Key R\&D Program of China No. 2023YFA1607903, the NSFC (grants 12273089, 12073023, 12233007, 12361131579, 12347103), and the science research grants from the China Manned Space Project with No. CMSCSST-2021-A06, and the Fundamental Research Fund for Chinese Central Universities.

\bibliography{CLAGN}
\bibliographystyle{aasjournal}

\end{document}